\documentclass[prd,showpacs,showkeys,nofootinbib,floatfix,eqsecnum,
               fleqn,preprint,12pt,tightenlines]{revtex4-1} 

\usepackage{amsmath,amssymb,revsymb,graphicx,dcolumn}
\usepackage{marginnote}

\newcommand{\beq}{\begin{equation}}
\newcommand{\eeq}{\end{equation}}
\newcommand{\beqa}{\begin{eqnarray}}
\newcommand{\eeqa}{\end{eqnarray}}
\newcommand{\bsubeqs}{\begin{subequations}}
\newcommand{\esubeqs}{\end{subequations}}

\newcommand{\christoffel}[3]{\Gamma^{#1}_{\hphantom{#1}#2#3}}

\begin{document}


\title{Regularized big bang singularity: Geodesic congruences}%
\vspace*{1mm}

\author{Z.L. Wang}
\email{ziliang.wang@just.edu.cn}
\affiliation{Department of Physics, School of Science, \\
Jiangsu University of Science and Technology, Zhenjiang, 212003, China\\}

\begin{abstract}
\vspace*{1mm}\noindent
We investigate a particular regularization of big bang singularity, which remains within the domain of $4-$dimensional general relativity but allowing for degenerate metrics.  We study the geodesics and geodesic congruences in the modified Friedmann–Lemaître–Robertson–Walker universe. In particular, we calculate the expansion of timelike and null geodesic congruences. Based on these results, we also briefly discuss the cosmological singularity theorems.

\vspace*{-0mm}
\end{abstract}


\maketitle

\section{Introduction}
\label{sec:Intro}
The expanding universe could be described by the Friedmann solution \cite{Friedman:1922kd,Friedmann:1924bb} of Einstein’s gravitational field equation.  However, this solution has a big bang singularity with divergent energy density. Also, it possesses incomplete geodesics according to the Hawking and Hawking-Penrose cosmological singularity theorems \cite{Hawking:1965mf,Hawking1967,Hawking1970}.

Recently, a particular regularization of the Friedmann big bang singularity has been proposed by Klinkhamer in Ref.~\cite{Klinkhamer2019}. This regularization is obtained within the realm of four-dimensional general relativity but allowing for degenerate metrics. The Friedmann big bang singularity is replaced by a three-dimensional  defect of spacetime with topology $\mathbb{R}^3$. Physical quantities, such as matter energy density and Ricci curvature scalar, are finite in the modified  Friedmann–Lemaître–Robertson–Walker (FLRW) universe.  In fact, the regularized big bang singularity can give rise to a nonsingular bouncing cosmology \cite{KlinkhamerWang2019-preprint-v3,Klinkhamer:2019gee,Klinkhamer:2019ycb}.  Potential experimental signatures of the nonsingular bounce are discussed by Klinkhamer and the present author \cite{KlinkhamerWang2019-preprint-v3}. Also, the work of Ref.~\cite{Klinkhamer:2019gee} has shown that the nonsingular bounce is stable under linear cosmological perturbations of the metric and matter. For more physical analysis of the nonsingular bouncing cosmology, see Ref.~\cite{Battista:2020lqv}.

The aim of the present paper is to study the geodesics and geodesic congruences in the background of the modified FLRW universe. In particular, we focus on the expansion of geodesic congruence, which is divergent at the Friedmann big bang singularity.

The outline of this paper is as follows. In Sec.~\ref{Regularized big bang singularity}, we review a particular regularization of the big bang singularity and give the solution for cosmic scale factor in the modified FLRW universe.  In Sec.~\ref{sec:Geodesics in the modified FLRW universe}, we study the geodesics in the modified FLRW universe and present the solution for timelike geodesics.  Subsequently, in Sec.~\ref{sec:Geodesic Congruences}, we investigate three kinds of geodesic congruences. Then, we discuss the Hawking and Hawking-Penrose cosmological singularity theorems. A brief summary is given in Sec.~\ref{Conclusion and Discussion}.

\section{Regularized big bang singularity}
\label{Regularized big bang singularity}

The particular regularization of the Friedmann big bang singularity is based on the following \emph{Ansatz} for metric \cite{Klinkhamer2019,KlinkhamerWang2019-preprint-v3}:
\bsubeqs\label{eq:modified-RW}
\begin{align}\label{eq:modified-RW-1}
d s^2 \Big|_\text{RWK} &\equiv g_{\mu \nu} (x) dx^{\mu} dx^{\nu}\Big|_\text{RWK}=-\frac{T^2}{b^2+T^2}dT^2 +a^2 (T)\delta _{ij} dx^i dx^j \,,\\ \label{eq:modified-RW-2}
b &>0\,,\\
T &\in (-\infty, \infty)\,,\\
x^i &\in (-\infty, \infty)\,,
\end{align}
\esubeqs
where we set $c=1$ and let the spatial indices $i,j$ run over $\{ 1, 2, 3\}$.

In fact, \eqref{eq:modified-RW-1} is a modified version of the spatially flat  Robertson–Walker (RW)  metric. First, it  gives  the standard spatially flat RW metric when $T \neq 0$ if $b = 0$. Second, for a nonvanishing parameter $b$ [as required by \eqref{eq:modified-RW-2}], we could define
\beq  \label{eq:def_t}
t (T)=
 \begin{cases}
    +\sqrt{b^2+T^2}\,, &  \text{for} \;T\geq 0 \,, \\
     -\sqrt{b^2+T^2}\,,  & \text{for} \; T\leq 0 \,.
 \end{cases} 
 \eeq
 Then, \eqref{eq:modified-RW-1}  can be written in a standard spatially flat RW metric form:
 \bsubeqs
 \begin{align} \label{eq:RW-1}
 ds^2 \Big| _\text{RWK} &= -dt^2+ \tilde{a} ^2(t) \delta _{ij}dx^i dx^j\,,\\
t&\in (-\infty,-b]\cup [b,+\infty)\,.
 \end{align}
\esubeqs

Several remarks are in order. First, $t$, as a function of the cosmic time $T$, is multivalued at $T=0$ ($t=-b$ and $t=b$ correspond to the single point $T = 0$), which leads to the fact that the differential structure of the metric \eqref{eq:modified-RW-1} is different from the one of the metric \eqref{eq:RW-1}. See Refs.~\cite{Klinkhamer2019,KlinkhamerWang2019-preprint-v3} for more related discussions.

Second,  the metric from \eqref{eq:modified-RW-1} is degenerate: $\text{det}\, g_{\mu \nu} =0$ at $T=0$, and the corresponding $T = 0$ (~$t= \pm b$) spacetime slice is a three-dimensional spacetime defect with characteristic length $b$. The terminology ``spacetime defect" is chosen to emphasize the
analogy with a defect in a crystal (Supposing that we cool a liquid rapidly, then the resulting crystal might be
imperfect, containing crystallographic defects. In a similar way, a spacetime defect might be a remnant  when classical spacetime emerges from some form of ``quantum phase".) See Refs.~\cite{Klinkhamer2019,Klinkhamer:2018xot,Klinkhamer:2019ocj} for further discussion on the spacetime defect and Ref.~\cite{Horowitz:1990qb} for a discussion on mathematical aspects of degenerate metrics. 

Third, the coordinate $t$ in  \eqref{eq:RW-1} is the proper time of (future-directed) co-moving observers. So, \eqref{eq:def_t} also represents the relation between the coordinate $T$ and the proper time of co-moving observers. We stress that the defect mentioned in the second remark actually appears in the direction of (proper) time. The possible origin of this defect will be discussed in Sec.~\ref{Conclusion and Discussion}.  For a similar type but a space defect, see Ref.~\cite{Klinkhamer:2018xot} and references therein.

For convenience, we will call the modified spatially flat RW metric as Robertson-Walker-Klinkhamer (RWK) metric in the remainder of this paper. Similarly, the corresponding modified FLRW universe will be called FLRWK universe.

With the metric \eqref{eq:modified-RW} and taking the energy-momentum tensor of a homogeneous perfect fluid [with energy density $\rho(T)$, pressure $P(T)$, and a constant equation-of-state parameter $w$], the Einstein equation leads to the following modified spatially flat Friedmann equations:
\bsubeqs\label{eq:modified-Friedman eq}
\begin{align}\label{eq:modified-Friedman eq-1}
&\left(1+\frac{b^2}{T^2}\right)\,\left(\frac{1}{a(T)}\,\frac{d a(T)}{dT}\right)^2 = \frac{8 \pi G_{N}}{3} \,\rho (T)\,,\\ \label{eq:modified-Friedman eq-2}
&\frac{b^2+T^2}{T^2}\left[\frac{1}{a(T)}\, \frac{d^2 a(T)}{d T^2} +\frac{1}{2}\,\left(\frac{1}{a(T)}\,\frac{da(T)}{dT}\right)^2\right]-\frac{b^2}{T^3}\,\frac{1}{a(T)}\,\frac{da(T)}{dT}=-4\pi G_{N} P(T)\,, \\
&\frac{d}{da} \left(a^3 \rho(a) \right)+ 3\,a^2\,P=0\,,\\
&\frac{P(T)}{\rho(T)}=w=\text{const}\,,
\end{align}
\esubeqs
where $G_N$ is Newton’s gravitational coupling constant. 

The modified Friedmann equations \eqref{eq:modified-Friedman eq-1} and \eqref{eq:modified-Friedman eq-2} are singular  differential equations (the singularities appear at $T = 0$) but they have a nonsingular solution, which will be given shortly. For comparison, 
remind that the standard Friedmann equations are nonsingular differential equations with a singular solution (the singularity is called the big bang singularity.) For more discussions on the mathematical structure of the modified Friedmann equations, see Refs.~\cite{Klinkhamer2019,Klinkhamer:2019gee}.

In general, the solution for $a(T)$ could be even or odd in $T$~\cite{Klinkhamer2019}. The $T-$odd solution could be of interest for a CPT-symmetric universe~\cite{Boyle:2018tzc}. The $T-$even solution, with positive definite cosmic scale factor, naturally gives a nonsingular bouncing universe~\cite{KlinkhamerWang2019-preprint-v3,Klinkhamer:2019gee}.   Energy density and  curvature scalars (Kretschmann curvature scalar and Ricci curvature scalar) are found to be \emph{finite} at $T=0$ \cite{Klinkhamer2019,KlinkhamerWang2019-preprint-v3,Klinkhamer:2019gee} for the FLRWK universe.

For a radiation-dominated universe ($w=1/3$) and a matter-dominated universe ($w=0$), the $T-$even solutions for $a(T)$ read~\cite{Klinkhamer2019,KlinkhamerWang2019-preprint-v3}
\bsubeqs\label{eq:a-FLRW}
\begin{align}\label{eq:a-radiation}
a (T) \Big|_\text{FLRWK}^\text{(w=1/3)} &=\sqrt[4]{\frac{b^2+T^2}{b^2+T_0 ^2}}\,,\\[2mm] 
\label{eq:a-matter}
a(T)\Big|_\text{FLRWK}^\text{(w=0)} &= \sqrt[3]{\frac{b^2+T^2}{b^2+T_0 ^2}}\,,
\end{align}
\esubeqs
with normalization $a(T_0) = 1$ for $T_0 > 0$.

The odd solutions for $a(T)$ are given by the right-hand side of \eqref{eq:a-FLRW} for $T>0$ and the same with an overall minus sign for $T<0$. The focus of this paper is on the $T-$even solutions, but we will see that the conclusions based on the $T-$even solutions also apply to the $T-$odd solutions for $a(T)$.

\section{Geodesics in the FLRWK universe}
\label{sec:Geodesics in the modified FLRW universe}
At the beginning of this section, we will show that, for the RWK metric \eqref{eq:modified-RW}, timelike and null geodesics are \emph{all} straight lines. Specifically, particles will travel on straight lines in the coordinate system $\{x^0 \equiv T,\,  x^{1},\, x^{2},\, x^{3}\}$.

Notice that the geodesic equation can be written as
\beq\label{eq:geodesic-eq-u}
\frac{d U_{\rho}}{d \lambda} - \frac{1}{2} \, \frac{\partial g_{\nu \beta}}{\partial x^{\rho}} U^{\nu} U^{\beta}=0\,,
\eeq
with $\lambda$ being the proper time for massive particle or the affine parameter for massless particle. (Note that the proper time for a massless particle is not well defined; the parameter $\lambda$ should be understood as the time told by some other freely falling clock.)

$U^{\mu}$ in \eqref{eq:geodesic-eq-u} is defined by
\beq\label{eq:4-velocity-definition}
U^{\mu} \equiv \frac{d x^{\mu}}{d \lambda}\,,
\eeq 
which is the four-velocity vector for a massive particle or energy-momentum four-vector for a massless particle. 

Recall that 
\bsubeqs\label{eq:g00_and_gij}
\begin{align}
g_{00}&=\frac{-T^2}{T^2 + b^2} \,, \\
g_{ij}&=a^2 (T) \delta _{ij}\,,
\end{align}
\esubeqs
which are independent of spatial coordinates. 

With \eqref{eq:g00_and_gij},  we could obtain from the geodesic equation \eqref{eq:geodesic-eq-u} that
\beq
\frac{d U_i}{d \lambda}=0\,,
\eeq 
i.e., spatial components of $U_{\mu}$ are constants along the geodesic in the coordinate system $\{T,\,  x^{1},\, x^{2},\, x^{3}\}$. For convenience, we write these constants as
\beq\label{eq:U123}
U_1 \equiv  c_1 \,\,,\,\,\,\,
U_2 \equiv  c_2\,\,,\,\,\,\, U_3 \equiv c_3\,.
\eeq
From the definition of $U^i$, we have
\bsubeqs
\begin{align}
\frac{d x^1}{d \lambda}& =\frac{c_1}{a^2(T)} \,, \\
\frac{d x^2}{d \lambda}& =\frac{c_2}{a^2(T)} \,, \\
\frac{d x^3}{d \lambda}& =\frac{c_3}{a^2(T)} \,,
\end{align}
\esubeqs
from which we can get
\beq\label{eq:app0-line-condition1}
\frac{dx^i}{dx^j} =\frac{dx^i /d\lambda}{dx^j / d \lambda} = \frac{c_i}{c_j}\,.
\eeq
\bsubeqs
From \eqref{eq:app0-line-condition1}, we could obtain the following parametric representation of a straight line in 3-space
\begin{align}\label{eq:line}
x^1 &= x^1 \,,\\
x^2 &= \frac{c_2}{c_1} x^1 + b_2   \,,\\
x^2 &= \frac{c_3}{c_1} x^1 + b_3  \,,
\end{align}
\esubeqs
with $x^1$ being the parameter and $b_{2,\,3}$ real constants.

Since particles travel on straight lines in the coordinate system $\{T,\,  x^{1},\, x^{2},\, x^{3}\}$, without loss of generality, we can consider geodesics that start at $T=T_1 <0$ and end at $T=T_0 >0$, while moving in the $x^1 \equiv X$ direction. So, we take $c_2 =c_3 =0$ and $c_1 >0$ in \eqref{eq:U123}. 

Notice that 
\beq\label{eq:dx/dt}
\frac{d X}{d T} = \frac{dX/d\lambda}{dT/d\lambda}= \frac{U^1}{U^0}\,,
\eeq
and the normalization
\beq
g_{\mu \nu} U^{\mu} U^{\nu} =N\,,
\eeq
with $N=0$ for massless particles and $N=-1$ for massive particles. Then, we have 
\beq\label{eq:u0 -v}
(U^{0})^2 = \left( -N+ \frac{c_1 ^2}{a^2}\right)\, \frac{b^2+T^2}{T^2}\,.
\eeq 
Taking into account \eqref{eq:u0 -v}, \eqref{eq:dx/dt} gives
\beq\label{eq:ode-X-T}
dX = \frac{c_1/a^2}{\sqrt{-N+c_1 ^2/a^2}} \, \sqrt{\frac{T^2}{b^2+T^2}}\,dT\,. 
\eeq

For null geodesic $N=0$, \eqref{eq:ode-X-T} reduces to
\beq\label{eq:eom-x}
dX = \sqrt{\frac{T^2}{a^2(T) (b^2+T^2)}}\,dT\,,
\eeq
which agrees with Eq.~(3.1) in Ref.~\cite{KlinkhamerWang2019-preprint-v3}. The solution for null geodesics has been derived in Ref.~\cite{KlinkhamerWang2019-preprint-v3}, so we will focus on timelike geodesics.

\begin{figure}[t]
\vspace*{0mm}
\begin{center}
\includegraphics[width=0.8\textwidth]{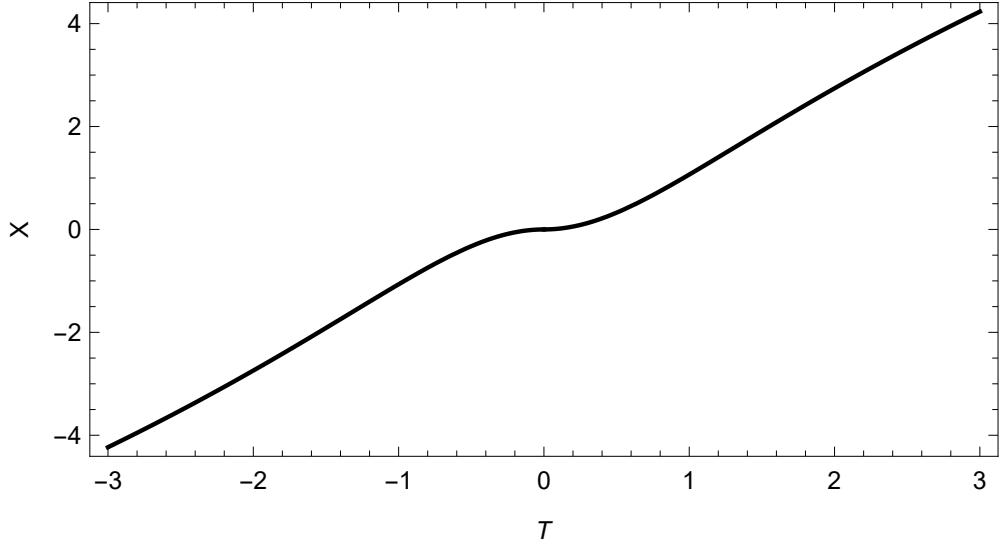}
\end{center}
\vspace*{-5mm}
\caption{Timelike geodesic \eqref{eq:timelike-geodesic} with $b=1$, $T_{0}=4\,\sqrt{5}$, $c_1=1$, and $c_4=-18 \tanh ^{-1}\left(1/\sqrt{10}\right) =-c_5$.}
\label{fig:timelike-geodesic}
\end{figure}

For radiation-dominated universe, the $T$-even solution for $a(T)$ is given by~\eqref{eq:a-radiation} (Remark that, since \eqref{eq:eom-x} depends on $a^2 (T)$, the solution for $X(T)$ will be the same for $T-$odd and $T-$even solution of $a(T)$.) In this case, the solution for timelike geodesics  is as follows:
\beq\label{eq:timelike-geodesic}
X(T) =
\begin{cases}
+\,2\,c_1\,F(T)\,\frac{ \sqrt{\sqrt{\frac{b^2+T^2}{b^2+T_0 ^2}}+c_1^2} }{\sqrt{\frac{1}{b^2+T^2}} \left(\frac{b^2+T^2}{b^2+T_0 ^2}\right)^{3/4} \sqrt{\frac{c_1^2}{\sqrt{\frac{b^2+T^2}{b^2+ T_0 ^2}}}+1}}+c_4 \,,
&\;\;\text{for}\;\;T > 0 \,,
 \\[16mm]
-\,2\,c_1\,F(T)\,\frac{\sqrt{\sqrt{\frac{b^2+T^2}{b^2+T_0 ^2}}+c_1^2}}{\sqrt{\frac{1}{b^2+T^2}} \left(\frac{b^2+T^2}{b^2+T_0 ^2}\right)^{3/4} \sqrt{\frac{c_1^2}{\sqrt{\frac{b^2+T^2}{b^2+ T_0 ^2}}}+1}}+c_5\,,
&\;\;\text{for}\;\;T \leq 0 \,,
 \end{cases}
\eeq
where
\beq
F(T) = \tanh ^{-1}\left(\frac{\sqrt[4]{\frac{b^2+T^2}{b^2+T_0^2}}}{\sqrt{\sqrt{\frac{b^2+T^2}{b^2+T_0^2}}\,+c_1^2}}\right)\,.
\eeq
 In \eqref{eq:timelike-geodesic}, $c_4$ is an arbitrary real constant and 
\beq
c_5 = 4b\,\frac{c_1 \, F(0)\, \sqrt{\sqrt{\frac{b^2}{b^2+T_0 ^2}}+c_1^2}}{\left(\frac{b^2}{b^2+T_0 ^2}\right)^{3/4} \sqrt{\frac{c_1^2}{\sqrt{\frac{b^2}{b^2+ T_0 ^2}}}+1}} + c_4 \,.
\eeq  
A plot of timelike geodesic is given in Fig.~\ref{fig:timelike-geodesic}.

We conclude that particles can travel from the prebounce phase to the postbounce phase which generalizes the conclusion in Sec.III.A of Ref.~\cite{KlinkhamerWang2019-preprint-v3}.

\section{Geodesic congruences in the FLRWK universe} 
\label{sec:Geodesic Congruences}
In this section, we will study timelike and null geodesic congruences in the  FLRWK universe and discuss the Hawking and Penrose singularity theorems.

\subsection{Timelike geodesic congruences}
\label{sec:Timelike geodesic congruences}
Let us start with some definitions. Consider a spacetime manifold ($M, g_{\mu \nu}$) and an open subset $O\subset M$. A geodesic congruence in $O$ is a family of curves such that each point in $O$ lies on one and only one geodesic of this family~\cite{wald1884general}.

Now, let $\xi^{\mu}$ be the tangent vector field of a geodesic congruence. Then, the behavior of the congruence can be described by  the the expansion $\theta$, the shear $\sigma _{\mu \nu}$ and the twist $\omega _{\mu \nu}$.  For timelike geodesic congruences, they are defined as~\cite{wald1884general}
\bsubeqs
\begin{align}\label{eq:expansion shear and twist}
\theta &\equiv B^{\mu \nu} h_{\mu \nu} \,,\\
\sigma _{\mu \nu}&\equiv \frac{1}{2}(B_{\mu \nu}+B_{\nu \mu})-\frac{1}{3}
\theta \, h_{\mu \nu} \,,\\
\omega _{\mu \nu} &\equiv \frac{1}{2} (B_{\mu \nu}-B_{\nu \mu})\,,
\end{align}
\esubeqs
where
\bsubeqs
\begin{align}
B_{\mu \nu} &\equiv \nabla _{\nu} \xi _{\mu}\,,\\
h_{\mu \nu} &\equiv g_{\mu \nu} +\xi _{\mu} \xi_{\nu}\,.
\end{align}
\esubeqs
Since $B_{\mu \nu}$ is ``spatial", i.e.,
\beq
B_{\mu \nu} \xi^{\mu}  = B_{\mu \nu} \xi^{\nu} =0\,,
\eeq
we have 
\beq \label{eq:theta}
\theta = B^{\mu \nu} g_{\mu \nu} = \nabla _{\mu} \xi^{\mu}\,.
\eeq

In a geodesic congruence, $\theta$ measures the expansion of nearby geodesics, i.e., $\theta >0$ means that the geodesics are diverging and $\theta <0$ means the geodesics are converging. $\sigma _{\mu \nu}$ measures the shear and $\omega_{\mu \nu}$ measure the rotation of nearby geodesics. 

Let us return to the FLRWK universe. The nonvanishing Christoffel symbols from the metric \eqref{eq:modified-RW} are given by
\bsubeqs\label{eq:christoffel}
\begin{align}
\christoffel{0}{0}{0}&=\frac{b^2}{T(T^2+b^2)}\,,\\
\christoffel{0}{i}{j}&=\frac{b^2+T^2}{T^2}a\,\dot{a}\,\delta _{ij}\,,\\
\christoffel{i}{0}{j}&=\frac{\dot{a}}{a}\,\delta_{ij}\,,
\end{align}
\esubeqs
where the overdot stands for differentiation with respect to $T$.

Now, we are able to study the geodesic congruences.

\subsubsection{Timelike geodesic congruences: Case one}
\label{Timelike geodesic congruences: Case one}
We start with the simplest case. Consider the geodesics given by the world lines of all co-moving observers in the FLRWK universe. A family of these curves is, of course, a congruence of timelike geodesics.

The vector field $\xi ^{\mu}$  tangent to the congruence is as follows:
\bsubeqs\label{eq:xi for FLRW}
\begin{align}
\xi ^0 (T)&= -\sqrt{\frac{b^2+T^2}{T^2}}\,,\\[2mm]
\xi ^i &=0\,.
\end{align}
\esubeqs
We emphasize that $\xi ^{\mu}$ is opposite to the four-velocity of the co-moving observers, as we are interested in the past-directed geodesic congruence.

Notice that, in this situation, $B_{\mu \nu}$ is equal to the extrinsic curvature of the constant$-T$ hypersurface and $h_{\mu \nu}$ is the induced metric of that hypersurface.   

The nonvanishing components of $B_{\mu \nu}$ and $h _{\mu \nu}$ are as follows:
\bsubeqs
\begin{align}
B_{i j} (T)&= -a\,\dot{a}\,\sqrt{\frac{b^2+T^2}{T^2}}\,\delta_{ij}\,,\\[2mm]
h_{ij}  (T)&= g_{ij}\,,
\end{align}
\esubeqs 
from which we can obtain
\bsubeqs
\begin{align}
\theta (T)&=-3\,\sqrt{\frac{b^2+T^2}{T^2}}\,\frac{\dot{a}}{a}\,,\label{eq:theta_case1}\\
\sigma _{\mu \nu} &=0 \,,\\
\omega _{\mu \nu} &=0\,,
\end{align}
\esubeqs
where the overdot stands again for differentiation with respect to $T$. Remark that, for $T-$odd and $T-$even solutions of the cosmic scale factor,  $\theta(T)$ will be the same.

Even though the shear  and twist are vanishing for the particular congruence discussed here, they can have nonvanishing value for more general congruences. Still, these two quantities are less interesting comparing with the expansion. So, we will focus on the expansion of geodesic congruences in the remainder of this paper.

For a radiation-dominated universe and a matter-dominated universe, we have 
\bsubeqs\label{eq:B_munu-FLRW-case1}
\begin{align}
B_{i j} (T) \Big|_\text{FLRWK}^\text{(case1;w=1/3)} &= -\frac{1}{2}\,\frac{T/|T|}{\sqrt{b^2+T_0^2}} \,\delta_{ij}\,,\\[2mm] 
B_{ij} (T)\Big|_\text{FLRWK}^\text{(case1;w=0)} &= -\frac{2}{3}\,\frac{T/|T|}{(b^2+T_0^2)^{2/3}}\,\sqrt[6]{b^2+T^2}\,\delta_{ij}\,,
\end{align}
\esubeqs
and expansion
\bsubeqs\label{eq:theta-FLRW-case1}
\begin{align}\label{eq:theta-FLRW-case1-1}
\theta (T) \Big|_\text{FLRWK}^\text{(case1;w=1/3)} &= -\frac{3}{2}\,\frac{T/|T|}{\sqrt{b^2+T^2}} \,,\\[2mm] \label{eq:theta-FLRW-case1-2}
\theta (T) \Big|_\text{FLRWK}^\text{(case1;w=0)} &= -2\,\frac{T/|T|}{\sqrt{b^2+T^2}}\,.
\end{align}
\esubeqs
Notice that, in both \eqref{eq:theta-FLRW-case1-1} and \eqref{eq:theta-FLRW-case1-2}, the expansion $\theta$ are negative for cosmic time  $T>0$.

With the solutions \eqref{eq:B_munu-FLRW-case1} and \eqref{eq:theta-FLRW-case1} in hand, we have four remarks in order.

First, as we mentioned before, the extrinsic curvature on the constant$-T$ hypersurface is equal to $B_{\mu \nu}$, i.e.,
\beq
K_{\nu \mu}  = B_{\mu \nu}\,.
\eeq
And the expansion of the timelike geodesic congruence is equal to the
trace of the extrinsic curvature on the constant$-T$ hypersurface, i.e.,
\beq
K(T) \equiv K^{\mu \nu} h_{\mu \nu} = \theta (T)\,.
\eeq

Second, for the FLRWK universe ($b \neq 0$), the expansion and the extrinsic curvature on constant$-T$ hypersurface are both discontinuous at $T = 0$. The discontinuities are a direct manifestation of the spacetime defect. 

Third, the expansion of the congruence for the standard FLRW universe is given by \eqref{eq:theta-FLRW-case1} with $b=0$. Then, we have  $\theta \to -\infty$ when  $T \to 0^{+}$. The singularity in the expansion $\theta$, which represents a singularity in the congruence, plays an important role in the proofs of the singularity theorems (see, e.g., Sec. IV of Ref.~\cite{wald1884general}).

Fourth, for the FLRWK universe ($b \neq 0$), $\theta (T)$ from \eqref{eq:theta-FLRW-case1} is always \emph{finite}. A finite $\theta$ is key for circumventing singularity theorems. More discussion on the singularity theorems will be given in Sec.~\ref{subsec:singularity theorem}.

So far, we have studied the geodesic congruence of the co-moving observers in the FLRWK universe. Actually, the conclusion, that the expansion $\theta$ has a finite discontinuity at $T=0$, can still hold for more general timelike and null geodesic congruences in the FLRWK universe.

\subsubsection{Timelike geodesic congruences: Case two}
\label{{Timelike geodesic congruences: Case two}}
Now, we consider a timelike geodesic congruence that each geodesic in the congruence has $c_1 >0$ and $c_2 =c_3=0$. In addition, all geodesics in the congruence have the same value of $c_1$. 

The vector field $\xi ^{\mu}$ tangent to the congruence is
\bsubeqs
\begin{align}
\xi ^0 (T)&= -\sqrt{\frac{b^2+T^2}{T^2}\,\frac{c_1 ^2 +a^2}{a^2}}\,,\\[2mm]
\xi ^1 (T) &=-\frac{c_1}{a^2}\,,\\
\xi ^2 &=\xi ^3 =0\,.
\end{align}
\esubeqs
The nonvanishing components of $B_{\mu \nu}$  are now given as follows:
\bsubeqs
\begin{align}
B_{00} (T)&= c_1^2\,\frac{\dot{a}}{a^3}\,\sqrt{\frac{T^2}{T^2+b^2}\,\frac{a^2}{a^2+c_1 ^2}}\,,\\[2mm]
B_{01}(T)&=B_{10}(T)=c_1 \frac{\dot{a}}{a} \,,\\[2mm]
B_{ij}(T)&=B_{ji}(T)=-\delta_{ij}\,a\,\dot{a}\,\sqrt{\frac{b^2+T^2}{T^2}\frac{c_1^2+a^2}{a^2}}\,.
\end{align}
\esubeqs 

Then, we can get the expansion of the congruence
\beq\label{eq:theta_case_2}
\theta (T)=
-\frac{\dot{a}}{a} \,\sqrt{\frac{b^2+T^2}{T^2}}\left(2\,\sqrt{\frac{c_1^2+a^2}{a^2}}+\sqrt{\frac{a^2}{c_1^2+a^2}}\right)\,,
\eeq
which reduces to \eqref{eq:theta_case1} if  $c_1=0$. Remark that,  $\theta(T)$ will be the same for $T-$odd and $T-$even solutions of $a(T)$.

The positive $c_1$ can be absorbed in the cosmic scale factor by replacing $a(T)/c_1$ by $a(T)$. Then \eqref{eq:theta_case_2} can be written as 
\beq
\theta (T)=-\frac{\dot{a}}{a} \,\sqrt{\frac{b^2+T^2}{T^2}}\left(2\,\sqrt{\frac{1+a^2}{a^2}}+\sqrt{\frac{a^2}{1+a^2}}\right)\,,
\eeq

For a radiation-dominated universe and a matter-dominated universe, we have 
\bsubeqs\label{eq:theta-FLRW}
\begin{align}\label{eq:theta-FLRW-1}
\theta (T) \Big|_\text{FLRWK}^\text{(case2;w=1/3)} &= -\frac{1}{2}\,\frac{T/|T|}{\sqrt{b^2+T^2}}\,\left(2\,\sqrt{\frac{1+a^2}{a^2}}+\sqrt{\frac{a^2}{1+a^2}}\right) \,,\\[2mm] \label{eq:theta-FLRW-2}
\theta (T) \Big|_\text{FLRWK}^\text{(case2;w=0)} &= -\frac{2}{3}\,\frac{T/|T|}{\sqrt{b^2+T^2}}\,\left(2\,\sqrt{\frac{1+a^2}{a^2}}+\sqrt{\frac{a^2}{1+a^2}}\right) \,,
\end{align}
\esubeqs
where $a(T)$ is given by \eqref{eq:a-radiation} and \eqref{eq:a-matter}, respectively. For nonsingular bouncing cosmology, the factor 
\beq\label{eq:prefactor}
\frac{T/|T|}{\sqrt{b^2+T^2}}
\eeq
allows for the expansion $\theta$ with a finite discontinuity at $T = 0$.
\subsection{Null geodesic congruences}
\label{sec:Null geodesic congruences}
Having discussed timelike geodesic congruence, we now turn to null geodesic congruences. 

For null geodesic congruences, we can still define $B_{\mu \nu}$ as
\beq
B_{\mu \nu}=\nabla _{\nu} k_{\mu}\,,
\eeq
where $k^{\mu}$ is the tangent null vector field. However, $h_{\mu \nu}$ is not unique for a given null geodesic congruence. (For a timelike geodesic congruence, $h_{\mu \nu}$ is unique once the tangent vector is determined.) See Chapter 2.4 of~\cite{poisson2004} for more discussion on null geodesic congruence.

Despite the nonuniqueness of $h_{\mu \nu}$, it can be proved that the expansion is still unique and given by~\cite{poisson2004}:
\beq\label{eq:theta_null}
\theta (T)= \nabla _{\mu} k^{\mu}\,.
\eeq

Similar to the second case of timelike geodesic congruence, we consider a null geodesic congruence that each geodesic in the congruence has $c_1 >0$ and $c_2 =c_3=0$. In addition, all geodesics in the congruence have the same value of $c_1$. 

Then, we have
\bsubeqs
\begin{align}
k^0 (T)&= - \sqrt{\frac{b^2+T^2}{T^2}\,\frac{c_1^2}{a^2}}\,,\\[2mm]
k^1(T)&=-\frac{c_1}{a^2}\,,\\
k^2&=k^3=0\,.
\end{align}
\esubeqs
Remark that, for null geodesic, $c_1$ cannot be $0$.

For this null geodesic congruence, the expansion is 
\beq
\theta(T)=-2\,\frac{\dot{a}}{a}\sqrt{\frac{b^2+T^2}{T^2}\frac{c_1^2}{a^2}}\,,
\eeq 
which can be written as
\beq
\theta(T)=-2\,\frac{\dot{a}}{a}\sqrt{\frac{b^2+T^2}{T^2}\frac{1}{a^2}}
\eeq
by rescaling the cosmic scale factor. Again, $\theta(T)$ is identical for $T-$odd and $T-$even solutions of the cosmic scale factor.

For a radiation-dominated universe and a matter-dominated universe, we have 
\bsubeqs\label{eq:theta-FLRW-null}
\begin{align}\label{eq:theta-FLRW-null-1}
\theta (T) \Big|_\text{FLRWK}^\text{(null;w=1/3)} &= -\frac{T/|T|}{\sqrt{b^2+T^2}}\,\sqrt[4]{\frac{b^2+T_0 ^2}{b^2+T^2}}\,,\\[2mm] \label{eq:theta-FLRW-null-2}
\theta (T) \Big|_\text{FLRWK}^\text{(null;w=0)} &= -\frac{4}{3}\,\frac{T/|T|}{\sqrt{b^2+T^2}}\,\sqrt[3]{\frac{b^2+T_0 ^2}{b^2+T^2}}\,.
\end{align}
\esubeqs
The factor \eqref{eq:prefactor} appears again and the expansion $\theta$ for the null geodesic congruence also has a finite discontinuity at $T = 0$.

\subsection{Singularity theorems}
\label{subsec:singularity theorem}

According to the calculation on geodesic congruences in  Sec.~\ref{sec:Timelike geodesic congruences} and \ref{sec:Null geodesic congruences},  there are now two scenarios:

\begin{itemize}
\item[1.]
In the standard FLRW universe, the expansion $\theta$ are singular at $T=0$. More precisely, $\theta \to -\infty$ when $T \to 0^{+}$.  For the explicit expressions of $\theta$, see \eqref{eq:theta-FLRW-case1}, \eqref{eq:theta-FLRW} and \eqref{eq:theta-FLRW-null} with $b=0$.
\item[2.]
In the FLRWK universe, the expansion $\theta$ are finite but discontinuous at $T=0$ (See \eqref{eq:theta-FLRW-case1}, \eqref{eq:theta-FLRW} and \eqref{eq:theta-FLRW-null} with $b\neq0$.)
\end{itemize}

These two scenarios lead to different results with regard to singularity theorems \cite{Hawking:1965mf,Hawking1967,Hawking1970}. In order to make a concrete comparison, we focus on the geodesic congruence of the co-moving observers in different scenarios. 

Consider a given constant $T=T_1 >0$ hypersurface. For the first scenario,  the ``point” $T=0 ^{+}$ is conjugate to  that hypersurface (For a spacelike hypersurface $\Sigma$ and a timelike geodesic congruence orthogonal to $\Sigma$, the sufficient and necessary condition for a point $p$ to be conjugate to $\Sigma$ is that the expansion of the congruence must go to $-\infty$ at point $p$. See Chapter.~9.3 of~\cite{wald1884general} for the proof of this statement.)  In general, the existence of conjugate points reveals the existence of extreme length curves.  In our case, the length of the past-directed co-moving observer's curve from $\Sigma$ has an upper bound. For radiation-dominated universe, the upper bound is $-{3}/{[2\,\theta (T_1)]} $ and for matter-dominated universe, the upper bound is $-{2}/{\theta (T_1)} $.  So, we have incomplete timelike geodesics, and the singularity theorems cannot be avoided.

 For the second scenario, the point conjugate to the constant $T_1$ hypersurface does not exist since the expansion of the congruence is always \emph{finite}. The length of the past-directed co-moving observer's curve has no upper bound. In this sense, the singularity theorem would be circumvented.

It is well known that the Raychaudhuri equation is of vital importance to singularity theorems. So let us now discuss the  Raychaudhuri equation in the background of RWK metric. Considering  the past-directed geodesic congruence of the co-moving observers, the Raychaudhuri equation is given by \cite{wald1884general}\footnote{We thank the referee for emphasizing the Raychaudhuri equation.} 
\beq\label{eq:Raychaudhuri equation}
\xi^{\mu}\nabla _{\mu} \theta=-\sqrt{\frac{T^2+b^2}{T^2}}\frac{d\theta}{dT} = -\frac{1}{3}\,\theta ^2-R_{ \mu \nu} \xi ^{\mu} \xi ^{\nu}\,,
\eeq
where $R_{\mu \nu}$ is the Ricci tensor. For a perfect fluid, we have 
\beq
R_{ \mu \nu} \xi ^{\mu} \xi ^{\nu} =4\pi G_{N} (\rho +3P)\,.
\eeq
For the matter content which satisfies the strong energy condition, $R_{ \mu \nu} \xi ^{\mu} \xi ^{\nu} $ is nonnegative. Moreover, for radiation-dominated and matter-dominated universe, $R_{ \mu \nu} \xi ^{\mu} \xi ^{\nu} $ is positive and we could obtain from \eqref{eq:Raychaudhuri equation} that 
\beq
\frac{d\theta}{dT}\geq 0\,,
\eeq
for past-directed geodesic congruence. The analysis on the Raychaudhuri equation agrees with our results in Sec.~\ref{Timelike geodesic congruences: Case one} (It can be checked that our solutions for $\theta$, i.e., \eqref{eq:theta-FLRW-case1}, satisfy the Raychaudhuri equation.) Remark that the expansion $\theta$ has continuous first-order derivative at $T = 0$ even though it is discontinuous at $T=0$. 


The regularized-big-bang model we studied in this paper  obeys the standard Einstein equation but has a vanishing determinant over the spacetime defect (mathematically, this defect is a three-dimensional submanifold of the spacetime manifold.) The existence of the spacetime defect with degenerate metrics is the key assumption for the model we studied in this paper. This assumption, of course, was not included in the proving of the Hawking-Penrose singularity theorems (Remind that the Hawking-Penrose singularity theorems is based on Einstein's general relativity, which assumes at the beginning that the determinant of the metric vanishes nowhere. For a historical discussion of standard general relativity, see Ref~\cite{Klinkhamer2019}.)

As the last part of this subsection, we would like to compare the nonsingular bouncing cosmology ($T-$even solution for the cosmic scale factor) discussed in this paper with other bouncing cosmologies. 

In the context of the standard general relativity, most bouncing cosmologies \cite{Novello:2008ra} in the literature require a violation of the strong energy condition. The violation of the strong energy condition \footnote{The violation of strong energy condition may lead to instabilities and  problems, as regards microcausality \cite{Novello:2008ra,Brandenberger:2016vhg}.}  can also lead to a finite expansion at the cosmic bounce and singularity theorems are avoided.

\section{Conclusions and Discussion}
\label{Conclusion and Discussion}

In the present paper, we have studied the geodesics and geodesic congruences in a modified FLRW universe, namely the FLRWK universe \cite{Klinkhamer2019,KlinkhamerWang2019-preprint-v3}.  We showed that all geodesics are straight lines in the RWK metric, just as in the standard RW metric. With this observation, we obtained the solution for geodesics in the FLRWK universe. The geodesic solution (Fig.~\ref{fig:timelike-geodesic}) indicates that particles can travel across the spacetime defect in the FLRWK universe (Another example of particles crossing a spacetime defect was studied in Ref.~\cite{Klinkhamer:2018ohw}. The Skyrmion spacetime defect in that reference could bring about a new type of gravitational lensing.)

The expansion of geodesic congruence can remain finite at $T=0$ for the FLRWK universe, while it is divergent at the big bang singularity for standard FLRW universe. For a geodesic congruence which starts at postbounce phase and goes back to the prebounce phase, a finite expansion along geodesics indicates the nonexistence of conjugate points, hence the length of the past-directed geodesic has no upper bound. In this sense, geodesics are extensible in the past direction for the FLRWK universe and the big bang singularity is regularized. 

Along with this regularization of big bang singularity, there is (finite) discontinuity at $T=0$ in the expansion of geodesic congruence. Taking the geodesic congruence of the co-moving observers as an example, the changes in the expansion at $T=0$ is given by $3/b$ for radiation-dominated universe. This discontinuity is a key manifestation of the three-dimensional spacetime defect with topology $\mathbb{R} ^3$ (More observables that could  reveal the presence of this spacetime defect were discussed in Refs.~\cite{KlinkhamerWang2019-preprint-v3,Battista:2020lqv}.)

Based on the calculations in Sec.~\ref{sec:Geodesic Congruences}, a finite expansion along the geodesics in the FLRWK universe may implies the circumvention of Hawking and Hawking-Penrose cosmological singularity theorems \cite{Hawking:1965mf,Hawking1967,Hawking1970}. However, due to the discontinuity in the expansion of geodesic congruence at the spacetime defect, it would be more appropriate to have the following interpretation: the singularity theorems are still valid in the FLRWK universe but the ``singularity" of these theorems corresponds to a spacetime defect with a local degenerate metric \cite{Klinkhamer:2021nub}.  This interpretation actually brings us to Hawking's question on the nature of the singularity \cite{Klinkhamer:2021nub}. We refer to the last paragraph in Sec.~3.3 of Ref.~\cite{Klinkhamer:2021nub} (and references therein) for a related discussion on the nature of the singularity.

The main intriguing task for the particular regularization of big bang singularity is to find the physical origin of the spacetime defect, which is also a crucial step to understand the nontrivial evolution of the geodesic congruences  discussed in this paper.  The spacetime defect with a degenerate metric may trace back to the underlying (unknown) theory of “quantum spacetime”. In loop quantum gravity \cite{Rovelli,Rovelli2014}, there does exist something like a “quantum of cosmic time” (cosmological evolution is discrete, see Sec.~8.1 of Ref.~\cite{Rovelli}), but the validity of the theory has not yet been established. It may very well be that the spacetime defect has its origin in string theory.  By comparing with string cosmology, it was found in Ref.~\cite{Klinkhamer:2019ocj} that the length scale of the spacetime defect could be related to the inverse of the string tension and may have the order of the Planck length. Moreover, recent research \cite{Klinkhamer:2020wct,Klinkhamer:2020xoi} on the large-N master field of the IIB matrix model~\cite{Ishibashi:1996xs,Aoki:1998bq} (a model which has been suggested as a formulation of nonperturbative type-IIB superstring theory) has shown the possibility to have a degenerate metric relevant to the regularized big bang.

\section*{Acknowledgements}

It is a pleasure to thank F.R.~Klinkhamer for informative discussions over the last years.
 


\end{document}